\documentclass[aps,pr,twocolumn,superscriptaddress,groupedaddress]{revtex4}  
\usepackage{graphicx}  
\usepackage{dcolumn}   
\usepackage{bm}        
\usepackage{amssymb}   
\usepackage{amsmath}
\usepackage{ascmac}
\usepackage{color}
\begin{document}

\title{Universality in Terms of The BECs Bloch-Sphere Manipulations”}
\input 
\author{Genji Fujii\\Department of Nuclear Engineering, Kyoto University, 6158540 Kyoto, Japan\\}
\date{\today}

\begin{abstract}
Quantum computing promises to outperform classical computing for certain computational tasks. One of the key concepts underlying this potential is universality. Universality has been extensively studied not only for qubits but also for higher-dimensional quantum systems, such as qutrits and qudits, which span Hilbert spaces of dimension greater than two. However, the concept of universality in Bose-Einstein condensates (BECs) qubit systems, which have been proposed relatively recently, remains insufficiently understood. In this work, we analyzed universality in BECs qubit systems. Our results clarify the theoretical aspects of universality in quantum computation within the class of ((N+1))-dimensional representations of SU(2), taking into account two distinct representations of the quantum states.
\end{abstract}

\maketitle


\section{introduction}
The universality of a quantum bit (qubit)\cite{1,2,3,4,5,6}, namely, the ability to implement arbitrary unitary time evolutions in a given Hilbert space, is one of the key indicators of the potential of quantum computers to outperform classical computers for certain computational problems\cite{7,8,9}. A gate set that provides universality in a quantum circuit is called a universal gate set. Similarly, quantum computation implemented using a universal gate set is referred to as universal quantum computation. Universal quantum computation can be viewed as a combination of gates that can move the Bloch vector to an arbitrary point on the Bloch sphere and gates that generate entanglement between two qubits. Examples of gate sets that enable universal quantum computation include the Hadamard, T, and controlled-Z (CZ) gates. In particular, it is well known that gates belonging to the class known as non-Clifford gates are required to move the Bloch vector to an arbitrary point on the Bloch sphere of a single qubit. Although quantum computational capabilities are enhanced by incorporating such non-Clifford gates, their implementation can conflict with transversality\cite{10,11} and may pose challenges for the realization of quantum error-correcting codes\cite{12,13}.\\
\indent The concept of universality has been extensively studied for qubits\cite{1,2,3,4,5,6}, which span a two-dimensional Hilbert space, and can be understood intuitively through the visual representation provided by the Bloch sphere. Indeed, the geometric analysis of the Bloch sphere has developed into a major area of research in quantum computation\cite{14,15,16,17,18,19,20}. However, this concept becomes nontrivial when the dimension of the Hilbert space exceeds two. Even in the simplest higher-dimensional case of a qutrit, the corresponding generalized Bloch sphere exhibits a highly intricate structure\cite{21,22,23,24}, which is one of the factors that makes the analysis of universality in quantum systems spanning Hilbert spaces of dimension three or higher challenging.\\
\indent The situation becomes even more complicated for Bose-Einstein condensates (BECs) qubits\cite{25}. BECs qubits can be represented by the Bloch sphere associated with SU(2), while, when the quantum state is interpreted in the Fock basis, it can also be regarded as a qudit-like quantum system spanning a higher-dimensional Hilbert space of dimension (N+1). Thus, for BECs qubits, analyzing the universality of two BECs qubits while taking entanglement into account requires the investigation of three distinct patterns of unitary time evolution arising from the combination of the qudit-like Fock-basis representation and the Bloch-sphere representation associated with SU(2).\\
\indent In this paper, we analyzed the universality of BECs qubits. We first identify some sets of the BECs qubits version of stabilizer operators and the states stabilized by them. Here, a stabilizer is defined as a set of operators that leave a given quantum state invariant. We then define some sets of the BECs qubits version of Clifford and non-Clifford gates and discuss the BECs qubits version of the Gottesman–Knill theorem\cite{26,27}, which is one of the central theorems in the theory of universal quantum computation. Finally, we present a stronger result concerning the universality of BECs qubits.\\
\indent As mentioned above, BECs qubits correspond, in terms of group representations, to the class of ((N+1))-dimensional representations of SU(2). Consequently, two distinct representations of the quantum states can be considered. A starting point for our analysis is the approach presented in Ref. \cite{28}, which investigates qudit universality by regarding pairs of two-level subspaces within a higher-dimensional Hilbert space as small Bloch spheres. In this approach, arbitrary unitary time evolutions within the subspaces of the higher-dimensional Hilbert space are considered, and these subspaces are successively combined to cover the entire Hilbert space. Furthermore, qudit universality can be established by introducing a single type of two-qudit gate.\\
\indent In the present work, we consider two complementary representations: small Bloch spheres spanned by subspaces of the higher-dimensional Hilbert space\cite{28}, represented in the Fock basis, and a large Bloch sphere of radius (N) associated with the SU(2) representation, represented in terms of BECs coherent states\cite{25}. Based on these two representations, we introduce the concept of Controlled-Bloch-Sphere Universality (CBSU). Note that, in this paper, our notion excludes cases in which the Bloch sphere is distorted due to errors.
\section{Gottesman-Knill theorem with BECs}
\subsection{A Stokes-stabilizer form}
In this sub section, we define some concepts for showing the BECs type Gottesman-Knill theorem. The $N$-particles BECs coherent state\cite{25}, we call this state BECs qubit in this paper, is defined as
\begin{equation}
|\alpha,\beta\rangle\rangle\equiv \frac{1}{N!}(\alpha a^{\dagger}+\beta b^{\dagger})^{N}|{\rm vac}\rangle,
\end{equation}
where $|{\rm vac}\rangle$ is a vaccum state, $a^{\dagger}$ and $b^{\dagger}$ are creation operators obeying commutation relations $[a,a^{\dagger}]=[b,b^{\dagger}]=1$, and $\alpha$ and $\beta$ are arbitrary complex numbers satisfying $|\alpha|^{2}+|\beta|^{2}=1$. Also, we introduce the Fock states:
\begin{equation}
\begin{split}
|k\rangle \equiv\frac{(a^{\dagger})^{k}(b^{\dagger})^{N-k}}{\sqrt{k!(N-k)!}}|{\rm vac}\rangle,
\end{split}
\end{equation}
where $k$ is an integer satisfying $0\leq k \leq N$. It is noteworthy that the states $|1,0\rangle\rangle=|N\rangle$ and $|0,1\rangle\rangle= |0\rangle$ can be treated as both qudit-like Fock states and BEC coherent states. In order to evolve unitary operation, we introduce Stokes operators. The Stokes operators are defined as
\begin{equation}
\begin{split}
S^{x}&=a^{\dagger}b+b^{\dagger}a,\\
S^{y}&=-ia^{\dagger}b+ib^{\dagger}a,\\
S^{z}&=a^{\dagger}a-b^{\dagger}b.\\
\end{split}
\end{equation}
\indent Also, we define an Identity operator, which is $S^{0}=I$. Furthermore, we define a Stokes-Hermite set $S_{H}$ and a Stokes-Unitary set $S_{U}$ as
\begin{gather}
S_{H}=\{\eta \frac{S^{\xi}}{N}|\eta=\pm1,\pm i, \xi =x,y,z\}.\\
S_{U}=\{\eta e^{-iS^{\xi}\pi/2}|\eta=\pm1,\pm i, \xi =x,y,z\}.
\end{gather}
\indent By choosing a suitable $\eta$, Stokes-Hermite operators become observable because they are Hermitian, but they are not unitary so they are not gate operators. In contrast to the Stokes-Hermite set, they are unitary, so they can describe quantum computing processes, but they are not Hermitian, so they are not observables. Stokes-Hermite operators stabilize some BECs qubits and Fock states:
\begin{equation}
\begin{split}
\frac{S^{x}}{N}|\frac{1}{\sqrt{2}},\frac{1}{\sqrt{2}}\rangle\rangle&=|\frac{1}{\sqrt{2}},\frac{1}{\sqrt{2}}\rangle\rangle,\\
\frac{S^{x}}{N}|\frac{1}{\sqrt{2}},\frac{-1}{\sqrt{2}}\rangle\rangle&=(-1)|\frac{1}{\sqrt{2}},\frac{-1}{\sqrt{2}}\rangle\rangle,\\
\frac{S^{z}}{N}|N\rangle&=|N\rangle,\\
\frac{S^{z}}{N}|0\rangle&=(-1)|0\rangle.\\
\end{split}
\end{equation}
\indent Also, the Stokes-Unitary operators stabilize some BECs qubitss and Fock states:
\begin{equation}
\begin{split}
e^{-iS^{x}\pi/2}|\frac{1}{\sqrt{2}},\frac{1}{\sqrt{2}}\rangle\rangle&=(-i)^{N}|\frac{1}{\sqrt{2}},\frac{1}{\sqrt{2}}\rangle\rangle,\\
e^{-iS^{x}\pi/2}|\frac{1}{\sqrt{2}},\frac{-1}{\sqrt{2}}\rangle\rangle&=(i)^{N}|\frac{1}{\sqrt{2}},\frac{-1}{\sqrt{2}}\rangle\rangle,\\
e^{-iS^{z}\pi/2}|N\rangle&=e^{-iN\pi/2}|N\rangle,\\
e^{-iS^{z}\pi/2}|0\rangle&=e^{iN\pi/2}|0\rangle.
\end{split}
\end{equation}
\indent In case of qubits, the Pauli operators, which are most fundamental, are both unitary and Hermitian. However, in case of BECs systems, the fundamental operators are classified as either Hermitian or unitary. This distinction complicates the situation, so special attention is required when introducing concepts such as the Gottesman-Knill theorem with BECs.
\subsection{Clifford and non-Clifford gates of BECs}
\indent Gottesman-Knill theorem, according to the expression in Ref. \cite{26,27}, says that ``{\it Any quantum computer performing only: a)
Clifford group gates, b) measurements of Pauli group operators, and c) Clifford
group operations conditioned on classical bits, which may be the results of earlier
measurements, can be perfectly simulated in polynomial time on a probabilistic
classical computer}.''\\
\indent Considering the BECs systems’ ability of calculations, it is worth revealing whether a BECs analog of the theorem exists. In other words, are there any conditions under which we can simulate a BECs system with a classical computer in polynomial time? In qubit systems, Clifford-gate-only circuits can be simulated by classical computers in polynomials time. There are several methods to demonstrate this, and one simple approach involves tracking stabilizer operators in the Heisenberg picture through Hamiltonian time evolution.\\
\indent Clifford gates possess the property of mapping Pauli operators to Pauli operators. In other words, if $C$ represents a Clifford gate and $P$ represents a Pauli operator, then $P' = C^{\dagger}PC$ is also a Pauli operator. In case of qubits, e.g, gates such as the Hadamard gate, Phase gate, CZ gate, and C-NOT gate are Clifford gates. In the context of BECs, e.g, Hadamard gate and Phase gate are defined in a way similar to gates for qubits:
\begin{equation}
\begin{split}
H&=e^{-iS^{y}3\pi/4},\\
S&=e^{-iS^{z}\pi/4}.
\end{split}
\end{equation}
In BECs systems, a gate that maps the Stokes-Hermite set to the Stokes-Hermite set is defined as a Clifford gate. By Campbell-Baker-Hausdorff formula, when applying rotation gates in the Bloch sphere to the Stokes-Hermite set, we get the following equations:
\begin{equation}
\begin{split}
e^{iS^{x}t}S^{x}e^{-iS^{x}t}&=S^{x},\\
e^{iS^{y}t}S^{x}e^{-iS^{y}t}&=\cos(2t)S^{x}+\sin(2t)S^{z},\\
e^{iS^{z}t}S^{x}e^{-iS^{z}t}&=\cos(2t)S^{x}-\sin(2t)S^{y},\\
e^{iS^{x}t}S^{y}e^{-iS^{x}t}&=\cos(2t)S^{y}-\sin(2t)S^{z},\\
e^{iS^{y}t}S^{y}e^{-iS^{y}t}&=S^{y},\\
e^{iS^{z}t}S^{y}e^{-iS^{z}t}&=\cos(2t)S^{y}+\sin(2t)S^{x},\\
e^{iS^{x}t}S^{z}e^{-iS^{x}t}&=\cos(2t)S^{z}+\sin(2t)S^{y},\\
e^{iS^{y}t}S^{z}e^{-iS^{y}t}&=\cos(2t)S^{z}-\sin(2t)S^{x},\\
e^{iS^{z}t}S^{z}e^{-iS^{z}t}&=S^{z}.
\end{split}
\end{equation}
By considering appropriate time evolution, it becomes evident that the Hadamard gate and Phase gate are Clifford gates.\\
\indent Next, let us define some non-Clifford gates in BECs systems. Particularly important gates are the $T$ gate and the $Z$-axis rotation gate $R_{z}(\varphi)$, defined as
\begin{equation}
\begin{split}
R_{z}(\varphi)&=e^{-iS^{z}\varphi},\\
T&=e^{-iS^{z}\pi/8}.
\end{split}
\end{equation}
\subsection{The Gottesman-Knill theorem analogy with BECs}
\indent We insist Gottesman-Knill theorem analogy for BECs systems. We insist that {\it Any quantum computer performing only: a)
Clifford group gates, b) measurements of Fock state basis, and c) Preparation of BECs qubit of coherent states, can be perfectly simulated in polynomial time on a probabilistic classical computer}.\\
\indent After the time evolution by Stokes-Hermite operators, the probability of obtaining $k$ by the measurement is
\begin{equation}
P_{k}=\langle k|C_{m}\cdot\cdot\cdot C_{1}|\frac{1}{\sqrt{2}},\frac{1}{\sqrt{2}}\rangle\rangle\langle\langle \frac{1}{\sqrt{2}},\frac{1}{\sqrt{2}}|C_{1}^{\dagger}\cdot\cdot\cdot C_{m}^{\dagger}|k\rangle,
\end{equation}
where each $C_{1}, \dots, C_{m}$ is the Clifford gates. Noting eq. (6), this expression can be written as
\begin{equation}
\begin{split}
&P_{k}\\
=&\langle k|C_{m}\cdots C_{1}(\frac{S^{x}}{N})C_{1}^{\dagger}\cdots C^{\dagger}_{m}C_{m}\cdots C_{1}|\frac{1}{\sqrt{2}},\frac{1}{\sqrt{2}}\rangle\rangle\\
\times&\langle\langle \frac{1}{\sqrt{2}},\frac{1}{\sqrt{2}}|C_{1}^{\dagger}\cdots C_{m}^{\dagger}C_{m}\cdots C_{1}(\frac{S^{x}}{N})C_{1}^{\dagger}\cdots C_{m}^{\dagger}|k\rangle.
\end{split}
\end{equation}
Each $C_{1}, \dots, C_{m}$ apply to the Stokes-Hermite operator respectively. So Clifford gates transfer a Stokes-Hermite operator to a Stokes-Hermite operator, this imply that simulation is possible in polynomial time on classical computers. However, when non-Clifford gates are introduced into the circuit, the situation changes. Non-Clifford gates duplicate the Stokes-Hermite operators, leading to an exponential increase in the number of permutations. Therefore, simulation becomes hard on classical computers.
\section{Controlled-Bloch-Sphere Universality}
In this section, we discuss the main result of this work, namely, Controlled-Bloch-Sphere Universality (CBSU). This concept refers to the ability to achieve universality through arbitrary movement of the Bloch vector on the Bloch sphere, together with the ability to arbitrarily shrink or expand the radius of the Bloch sphere. The latter capability, namely, the ability to arbitrarily shrink or expand the radius of the Bloch sphere, can be achieved using a single type of entangling gate. This is because an entangling gate of a single type can be placed between appropriate one-qubit gates in a quantum circuit, allowing the radius of the Bloch sphere to be arbitrarily reduced or increased on  ($0\leq \mathbf{r} \leq 1$), where $\mathbf{r}$ is the scaling factor defined later.\\
\indent On the other hand, for BECs qubits, we need to consider three cases. Specifically, we consider combinations of the large Bloch sphere of radius (N) associated with SU(2), which we refer to as the BECs Bloch sphere, and the small Bloch spheres associated with two levels embedded in subspaces of the ((N+1))-dimensional qudit-like Hilbert space, which we refer to as the Qudit-like Fock state representation. Thus, we need to consider three possible types of universality: \\
Case 1, Qudit-like Fock state and Qudit-like Fock state;\\
Case 2, BECs Bloch sphere and BECs Bloch sphere;\\
Case 3, BECs Bloch sphere and Qudit-like Fock state.
\subsection{Single BECs Bloch sphere}
Similar to conventional qubits, BECs qubits can be represented on a Bloch sphere. However, it is difficult to determine the radius of the Bloch sphere directly from the magnitude of the Bloch vector in the density-matrix representation. This is because the density-matrix representation obtained by simply replacing the Pauli operators with the Stokes operators, as in the qubit case, cannot be constructed. Therefore, the radius is instead obtained from the sum of the squares of the expectation values of the Stokes operators. The radius of the Bloch sphere for a pure state is
\begin{equation}
\langle S^{x} \rangle^{2}+\langle S^{y} \rangle^{2}+\langle S^{z} \rangle^{2}=N^{2},
\end{equation}
which coincides with the particle number N. There are two types of operations on the Bloch sphere: 1. one-qubit unitary operations, which move the Bloch vector on the sphere, and 2. entangling gates, which reduce or increase the radius of the Bloch sphere. \\
\subsection{Single qudit-like Fock state two-level sub-space Bloch sphere}
For the qudit-like Fock-state representation, the radius of the small Bloch sphere associated with two levels is determined as follows.\\
\indent Considering the projection operator onto the two-level subspace
\begin{equation}
\operatorname{span}\{{|j\rangle,|k\rangle}
\},
\end{equation}
as
\begin{equation}
P_{jk}=|j\rangle\langle j|+|k\rangle\langle k|.
\end{equation}
$1$. Projection: We project the density matrix onto the two-level subspace as$P_{jk}\rho P_{jk}.$
In general, this state is not normalized. Its trace is given by
$
\rho_{jj}+\rho_{kk},
$
which represents the probability of finding the system in the two-level subspace upon measurement.\\
$2$. Normalization: The normalized density matrix as
\begin{equation}
\rho_{jk}=\frac{P_{jk}\rho P_{jk}}
{Tr(P_{jk}\rho)}.
\end{equation}
It then satisfies$\operatorname{Tr}\rho_{jk}=1,$and thus regards as two-level density matrix.\\
$3$. Evaluation of expectation values using the Pauli operators: Within the two-level subspace, we define the Pauli operators as
\begin{equation}
\begin{split}
&X_{j,k}^{F}=|j\rangle\langle k|+
|k\rangle\langle j|,
\\
&Y_{j,k}^{F}=-i|j\rangle\langle k|
+i|k\rangle\langle j|,
\\
&Z_{j,k}^{F}=|j\rangle\langle j|
-|k\rangle\langle k|.
\end{split}
\end{equation}
\indent The corresponding expectation values are then given by
\begin{equation}
\begin{split}
&\langle X_{j,k}^{F}\rangle=\operatorname{Tr}
\left(
\rho_{jk}X_{j,k}^{F}
\right),\\
&\langle Y_{j,k}^{F}\rangle=\operatorname{Tr}
\left(
\rho_{jk}Y_{j,k}^{F}
\right),\\
&\langle Z_{j,k}^{F}\rangle=\operatorname{Tr}
\left(
\rho_{jk}Z_{j,k}^{F}
\right).
\end{split}
\end{equation}
\indent Therefore,
\begin{equation}
\langle X_{j,k}^{F}\rangle+\langle Y_{j,k}^{F}\rangle+\langle Z_{j,k}^{F}\rangle,
\end{equation}
is the Bloch Sphere radius associated with the two-level subspace.
\subsection{Scaling factor of entangled bipartite systems}
\indent In contrast, there is another criterion for determining the radius of a entangled BECs qubits and qudit-like fock states Bloch sphere. The radius of the Bloch vector can be also expressed as follows as an indicator of whether entanglement is present:
\begin{equation}
\begin{split}
&\rho_{1}=\operatorname{Tr}(\rho_{1,2}),\\
&\operatorname{Tr}(\rho_{1}^{2})=\frac{1+|\mathbf{r}|^{2}}{2},
\end{split}
\end{equation}
where, $\rho_{1,2}$ denotes the density matrix of an arbitrary entangled pair of BECs qubits. Note that, on the BECs coherent state side, this trace is an ordinary partial trace. On the qudit-like Fock-state side, the state must first be projected using the appropriate projection operator and then properly renormalized. This also holds for BECs qubits, for which the radius takes the value $\operatorname{Tr}(\rho_{1}^{2})<1$ when entanglement is present and  $\operatorname{Tr}(\rho_{1}^{2})=1$ when it is absent. Using these two criteria for determining the radius of the BECs Bloch sphere, we newly define the radius as $|\mathbf{r}|^2 N$, where $|\mathbf{r}|^2$ ($0\leq \mathbf{r} \leq 1$) is defined as the scaling factor. Also, this scaling factor cannot be consistently defined for a qudit when the entire Hilbert space is considered. However, it is well defined when restricted to two levels, as well as for BECs qubits. Note that the concept of this scaling factor arises only in bipartite quantum systems. Furthermore, for a pure two-qubit state, the reduced density matrices of the two subsystems always have the same eigenvalues, and therefore the relation $|\mathbf{r_{1}}|=|\mathbf{r_{2}}|$ always holds.\\
\indent A two-qubit gate is required to reduce the scaling factor of a given quantum state to zero. In other words, within the conventional framework of universality, it is sufficient to have at least one entangling gate chosen from an appropriate class of entangling operations.\\
\indent For qudits, sometimes universality can be discussed without relying on the Bloch-sphere representation\cite{29,30,31}. In contrast, for BECs coherent states, the Bloch-sphere representation plays a more central role than the description of states solely in terms of the Hilbert space. Consequently, the concept of CBSU, which incorporates the notion of a scaling factor, becomes particularly important.
\subsection{Case1:Qudit-like Fock states and Qudit-like Fock states}
\includegraphics[width=85mm]{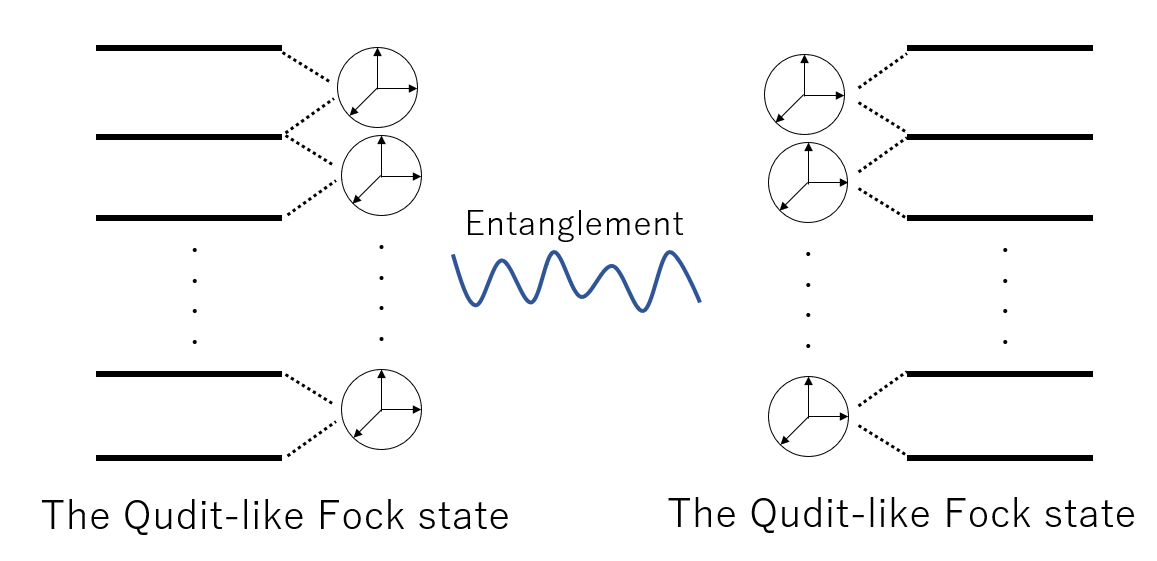}
\\
Fig.1. The Fock basis spans an (N+1)-dimensional Hilbert space consisting of the states $(|0\rangle,\ldots,|N\rangle)$. By regarding each pair of levels as a small Bloch sphere, entanglement between these subspaces can be introduced using a two-qudit gate. In particular, exact universality is achieved using the Hamiltonian given in Eq. (21) and Eq.(23). A bipartite system consisting of two such systems spans a Hilbert space of dimension $((N+1)\times(N+1))$.\\
\\
We discuss entanglement and universality  between qudit-like Fock states (Fig.1). In the context of BECs qubits, this corresponds to the class of entanglement viewed in Refs. \cite{32}, for example. The analysis of universality in this case has already been presented in Ref. \cite{28}, and here we briefly summarize its main idea.\\
For a (d)-level system, the authors consider the Hamiltonians
\begin{equation}
\begin{split}
H^x_{jk}
&=\hbar\Omega
\left(
|k\rangle\langle j|
+
|j\rangle\langle k|
\right),
\\
H^y_{jk}
&=\hbar\Omega
\left(
i|k\rangle\langle j|
-i|j\rangle\langle k|
\right).
\end{split}
\end{equation}
\indent In other words, these Hamiltonians control transitions between two levels,
\begin{equation}
|j\rangle\leftrightarrow |k\rangle.
\end{equation}
\indent These two gates enable arbitrary movement of the Bloch vector on the small Bloch sphere associated with any pair of levels. In addition, these gates enable transitions between arbitrary pairs of Fock-basis states. The authors further introduce the following two-qudit Hamiltonian,
\begin{equation}
H_{int}=-\hbar\Omega
|d-1,d-1\rangle
\langle d-1,d-1|.
\end{equation}
\indent This gate reduces the scaling factor between uniformly weighted qudits to zero. It is then shown that arbitrary single-qudit operations
and this two-qudit interaction
is sufficient to realize exact universality with qudits.\\
\indent Thus, arbitrary unitary transformations can be constructed from Hamiltonians that are allowed to act on the smalll Bloch spheres associated with two levels. Moreover, when arbitrary single-qudit operations are supplemented by the specific two-qudit interaction given above, exact multi-qudit universal quantum computation can be achieved.\\
\indent However, in the case of a qudit-like Fock state, it is not immediately clear whether applying a single entangling gate to the state in its original form can produce a state whose associated small Bloch sphere has zero radius. Since it is possible to transform the state into an equal-weight superposition using only sub-space single-qudit gates, one can instead apply the gate defined in $\sum_{k=0}^{N}\frac{|\alpha|^{k}|\beta|^{(N-k)}}{\sqrt{N!}}{ }_N C_k|k\rangle$ to this equal-weight superposition state, thereby generating a small Bloch sphere with zero radius.
\subsection{Case2:BECs Bloch sphere and BECs Bloch sphere}
\includegraphics[width=85mm]{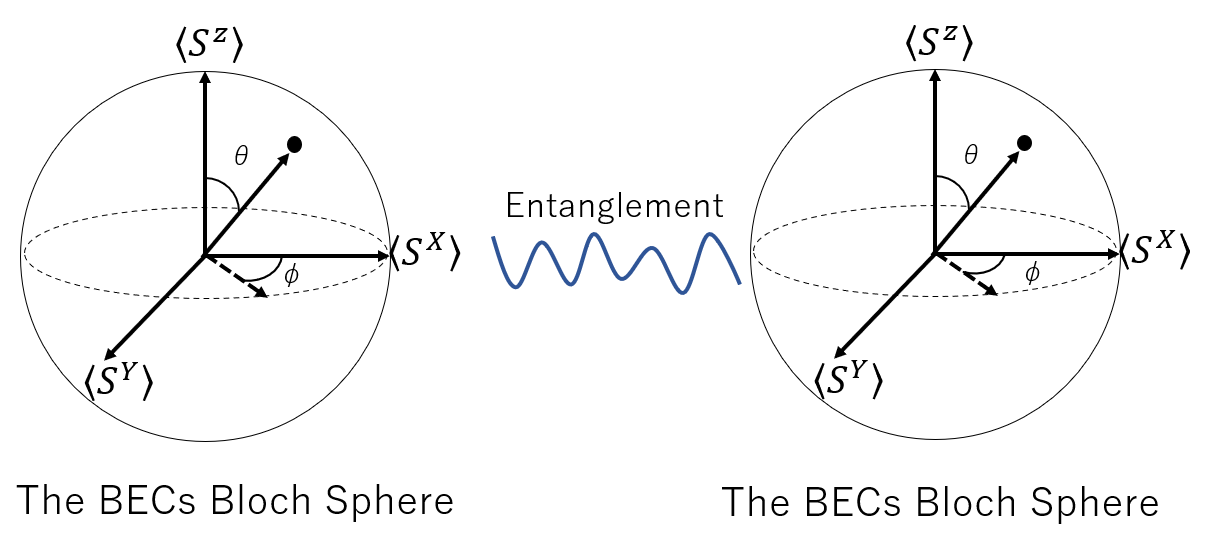}
\\
\\
Fig.2 The BECs Bloch sphere is described by a representation of SU(2) and can be regarded as a large Bloch sphere with radius (N). Entanglement between two BECs qubits is introduced through unitary time evolution generated by Stokes-operator interactions between them. An example of a maximally entangling state is given in Eq. (25).\\
\\
\\
We next discuss entanglement and universality between two BECs Bloch-sphere representations (Fig.2). In the context of BECs qubits, this corresponds to the class of entanglement viewed in Refs. \cite{33}, First, we note that arbitrary unitary time evolutions of a single BECs qubits can be implemented using a gate set that includes non-Clifford gates, and that the radius of the Bloch sphere can be calculated from Eq. (13). Furthermore, it is well known that, for a maximally entangled state, each subsystem is in a maximally mixed state, resulting in a Bloch-sphere radius of zero. Therefore, to realize the CBSU considered here, it is sufficient to have arbitrary single-BECs qubit unitary time evolutions together with a gate that can reduce the radius of the Bloch sphere to zero. Particularly, we consider the CZ gate proposed in Ref. \cite{33}, which provides the required operation.\\
\indent BECs coherent states can be represented on the BECs Bloch sphere only when they satisfy the following normalization condition for the probabilities. This condition as\\
\begin{equation}
\frac{|\beta|^{2N}}{N!}{ }_N C^{2}_0+\frac{|\alpha|^{2}|\beta|^{2(N-1)}}{N!}{ }_N C^{2}_1+\dots+\frac{|\alpha|^{2N}}{N!}{ }_N C^{2}_N=1.
\end{equation}
\indent This condition exhibits a remarkable structure among the states spanning the infinitely many possible higher-dimensional Hilbert spaces. Note that the condition is preserved when a BECs qubits are merely expanded in the Fock basis; however, it is generally violated when a projection is performed to investigate the small Bloch sphere associated with a qudit-like Fock state.\\
\indent In Ref. \cite{33}, the author define a maximally entangled graph state as follows.\\
\\
\begin{equation}
|G\rangle\rangle=|e\rangle|\frac{1}{\sqrt{2}},\frac{1}{\sqrt{2}}\rangle\rangle+|o\rangle|\frac{1}{\sqrt{2}},\frac{-1}{\sqrt{2}}\rangle\rangle.
\end{equation}
\indent At first glance, this state appears to belong to the class of qudit-like Fock states, since the subsystems are represented in the Fock basis. However, it can in fact be rewritten in terms of coherent states, as shown in 
\begin{equation}
\begin{split}
&|e\rangle=\frac{1}{2}(|\frac{1}{\sqrt{2}},\frac{1}{\sqrt{2}}\rangle\rangle+|\frac{1}{\sqrt{2}},\frac{-1}{\sqrt{2}}\rangle\rangle),\\
&|o\rangle=\frac{1}{2}(|\frac{1}{\sqrt{2}},\frac{1}{\sqrt{2}}\rangle\rangle-|\frac{1}{\sqrt{2}},\frac{-1}{\sqrt{2}}\rangle\rangle),
\end{split}
\end{equation}
and the expectation value calculated using Eq. (20) correctly yields the radius scaling factor corresponding to the BECs Bloch-sphere representation. In fact, when the expectation value is calculated after tracing out either of the subsystems of this graph state, the scaling factor ($\mathbf{r}$) is found to be zero.
\subsection{Case3:Qudit-like Fock states and BECs Bloch sphere}
\includegraphics[width=85mm]{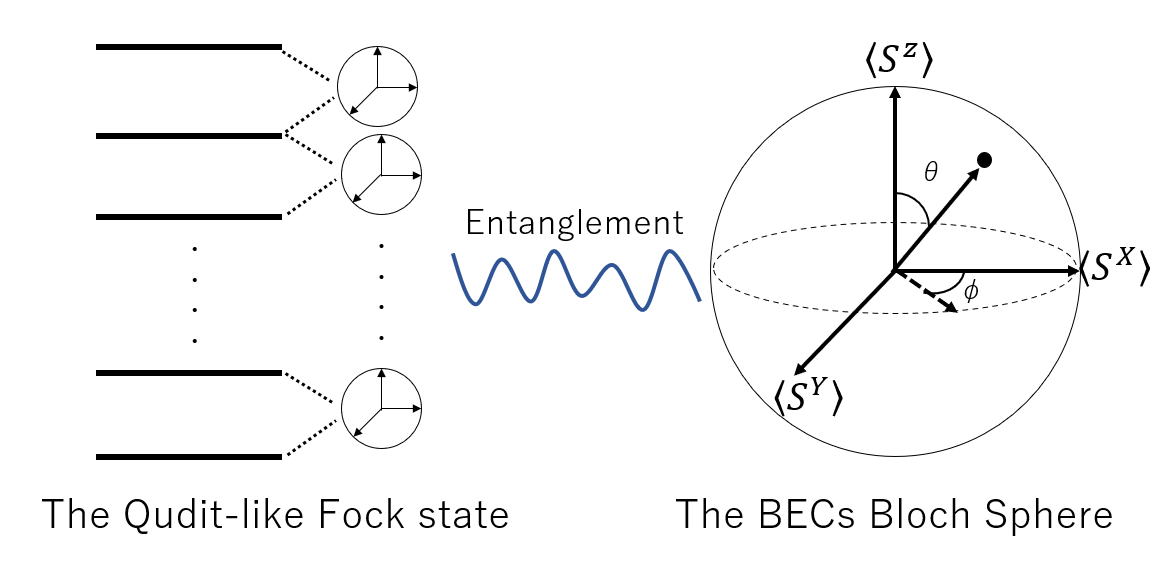}
\\
\\
Fig.3 Entanglement between a qudit-like Fock state and a BEC Bloch sphere. The two-quantum gate is given by a composite gate involving a qudit operation and Stokes operators.\\
\\
We finally discuss entanglement and universality between the qudit-like Fock state and the BECs Bloch sphere (Fig.3). We propose a Hamiltonian that reduces the radius of the Bloch sphere associated with the reduced density matrix to zero, as given in 
\begin{equation}
H=-|N\rangle\langle N|\otimes S^{z}.
\end{equation}
\indent Its action, in appropriate time evolution and entanglement between equal weight fock states and BECs coherent states, is described as 
\begin{equation}
\begin{split}
&\sum_{k=0}^{N}\frac{|k\rangle}{\sqrt{N+1}}|\frac{1}{\sqrt{2}},\frac{1}{\sqrt{2}}\rangle\rangle\\
\rightarrow&\sum_{k=0}^{N-1}\frac{|k\rangle}{\sqrt{N+1}}|\frac{1}{\sqrt{2}},\frac{1}{\sqrt{2}}\rangle\rangle+\frac{|N\rangle}{\sqrt{N+1}}|\frac{-1}{\sqrt{2}},\frac{1}{\sqrt{2}}\rangle\rangle.
\end{split}
\end{equation}\\
\indent It is then clear that the partial trace of this state yields the maximally mixed state, both in the two-level subspaces and in the BECs coherent state representation, resulting in the scaling factor of zero.
\section{Conclusion and Discussion}
In this paper, after presenting the BECs qubits version of the Gottesman–Knill theorem, we considered the three cases of the stronger concept of Controlled-Bloch-Sphere Universality (CBSU). For each case, we demonstrated that BECs qubits possess the characteristics of universality according to our proposed criterion.\\
\indent However, it is important to emphasize, as also noted in Ref. \cite{28}, that exact universality and computational efficiency are distinct concepts. Determining what information-processing properties are required for efficient computation, as well as clarifying how other independently developed concepts related to universality such as fault-tolerant quantum computation, the Eastin Knill theorem\cite{10}, and the relationship between transversality and universality\cite{11} fit into our framework, is left for future work.\\
\indent Another potentially interesting direction for theoretical development is the application of our framework to recent results concerning n-qubit systems \cite{34,35,36}. There may also be considerable scope for extending the theory using the Stokes-Hermite and Stokes-unitary concepts defined in our work.\\
\indent Also note that, in this paper, we have implicitly assumed that Fock states can be treated analogously to qudits in an ((N+1))-dimensional Hilbert space. Although this is theoretically possible, practical methods for manipulating such a high-dimensional Hilbert space in an experimentally accessible manner, analogous to qudit operations, still need to be developed. For example, how to implement Pauli gates in a qudit-like manner, how to distinguish these operations from those performed on the BECs qubit side, and how to realize entangling gates are important topics for future experimental work.
\input 

\end{document}